\documentclass[aps,prstab,twocolumn,superscriptaddress,groupedaddress,floatfix]{revtex4}  
\usepackage{tabularx}
\usepackage{graphicx}  
\usepackage{dcolumn}   
\usepackage{bm}        
\usepackage{amsmath}
\usepackage{amssymb}   
\usepackage[caption=false]{subfig}
\usepackage{bm} 
\usepackage{overpic}
\usepackage{natbib}
\usepackage[all,cmtip]{xy}

\newcommand{\bx}{{\bf x}}
\newcommand{\br}{{\bf r}}
\newcommand{\brp}{{\bf r}^{\prime}}
\newcommand{\eps}{\varepsilon}
\newcommand{\E}{{\bf E}}

\newcommand{\modepicwidth}{0.3\textwidth}
\newcommand{\abspicwidth}{0.365\textwidth}

\newcommand{\figpath}[1]{#1.pdf}

\begin{document}

\title{Origin and reduction of wakefields in photonic crystal
accelerator cavities}

\author{Carl A.~Bauer}
\author{Gregory R.~Werner}
\affiliation{University of Colorado at Boulder, Boulder, CO}
\author{John R.~Cary}
\affiliation{University of Colorado at Boulder, Boulder, CO}
\affiliation{Tech-X Corporation, Boulder, CO}

\date{\today}

\begin{abstract}

Photonic crystal (PhC) defect cavities that support an accelerating
mode tend to trap unwanted higher-order modes (HOMs) corresponding to
zero-group-velocity PhC lattice modes at the top of the bandgap.
The effect is explained quite generally by photonic band and
perturbation theoretical arguments.
Transverse wakefields resulting from this effect are observed in a
hybrid dielectric PhC accelerating cavity based on a triangular
lattice of sapphire rods. These wakefields are, on average, an order
of magnitude higher than those in the waveguide-damped Compact Linear
Collider (CLIC) copper cavities. The avoidance of
translational symmetry (and, thus, the bandgap concept) can dramatically
improve HOM damping in PhC-based structures.


\end{abstract}

\maketitle

\section{Introduction}

Photonic crystals (PhCs) have recently attracted interest from the
accelerator community
\cite{colby2002structure,
smith1997recent,kroll1993photonic,masullo2006study,
gennaro2009hybrid,gennaro2008mode,shapiro2004theoretical,
smirnova2005demo,marsh2011xband,bauer2008truncated,
cowan2008three,lin2001photonic}
for the following reasons:
\begin{enumerate}
\item PhCs enable the construction of accelerator cavities using
dielectric materials.
\item PhC cavities intrinsically
provide a wakefield damping mechanism.
\end{enumerate}
These features suggest an alternative to traditional (super)conducting
cavity design, and therefore \emph{could} result in higher gradients
\cite{thompson2008breakdown,bloem1974laser,du1994laser,power2004observation,jing2010progress},
lower power losses, and/or lower wakefields.

A simple argument from photonic band theory gives reason to believe
that PhC cavities will have low wakefields. It says that,
given a PhC with a bandgap, a defect cavity will
confine only one (ideally accelerating) mode to the defect; all other
higher-frequency modes will propagate through the crystal and
contribute minimally to the wakefields. This basic concept has led to
the design of many defect cavity PhC acceleration schemes, a sample of
which can be seen in the list of references above.

To date, only a handful of computational and experimental studies
dedicated to wakefield damping in PhC cavities have been performed.
In general, they have
shown that the Q-factors of higher-order modes (HOMs) 
are much lower than that of
the accelerating mode, indicating some wakefield damping
\cite{werner2009wakefields,marsh2009exp,jing2009observation}.
These studies were performed on
so-called hybrid PhC cavities (at GHz frequencies),
where ``hybrid'' indicates
the incorporation of both PhC {\it and} traditional
(metal disc-loaded waveguide) design concepts
(look ahead to Fig.~\ref{fig:tri4Cell} for an
example of a hybrid PhC cavity)
\cite{kroll1993photonic,smith1997recent,smirnova2005demo}.

This work looks in more detail at wakefield suppression in PhC
cavities and reveals subtleties that can undermine the benefits
implied by the simple bandgap argument. These subtleties were
uncovered in a thorough comparison between wakefield damping in
a hybrid PhC cavity based on a triangular lattice of sapphire rods
(Fig.~\ref{fig:tri4Cell}) and a cavity from the main linac of the
Compact Linear Collider (CLIC) which uses side-coupled
waveguides to damp HOMs (see Fig.~\ref{fig:clicCell})
\cite{wilson2000clic,grudiev2004newly,braun2008clic,grudiev2012clic}.

We find that, in the case of the hybrid
PhC cavity (with sapphire
rods) based on the triangular lattice, transverse wakefields are on
average higher than those in the CLIC waveguide-damped
cavity. A simple Fourier analysis of the transverse wake potential
compared with the triangular lattice band diagram shows that the
band edges (where the dispersion curves flatten) are highly correlated
with troublesome peaks in the transverse wake impedance. We attribute
this correlation to the slow rate at which low-group-velocity PhC modes
transport energy through the lattice.

We also find that PhC-based cavities optimized to confine the
accelerating mode with a minimal number of rods can
reduce wakefields compared to a PhC lattice cavity with similar
accelerating mode $Q$-factor.
The results of this paper show that the bandgap property of PhC-based
cavities is not enough to guarantee ideal wakefield suppression.
Given the configurability of PhCs (and our previous work in
Ref.~\cite{bauer2008truncated}), we suggest minimizing wakes through
rod-placement optimization.

The following two sections review waveguide and PhC wakefield damping
techniques. Section \ref{sec:results} describes our simulation methods
and results, comparing wakefields in the CLIC and lattice-based hybrid
dielectric PhC cavities.  Section \ref{sec:discuss} discusses the poor
damping found in the triangular-lattice-based PhC cavity using
photonic band theory. Finally, Section \ref{sec:opt} repeats the
simulations of Section \ref{sec:results} with an optimized cavity
(from Ref.~\cite{bauer2008truncated}) and compares these results with
CLIC. Appendix \ref{app:absorb} analyzes
the performance of our numerical absorbers used in wakefield
simulations.  Appendix \ref{app:accMode} discusses accelerating mode
figures of merit (peak surface fields, accelerating efficiency, etc.)
for each cavity type considered in this work.

\section{Waveguide damping and CLIC}
\label{sec:clicWg}

\begin{figure}[htbp]
\centering
\includegraphics[width=0.35\textwidth]{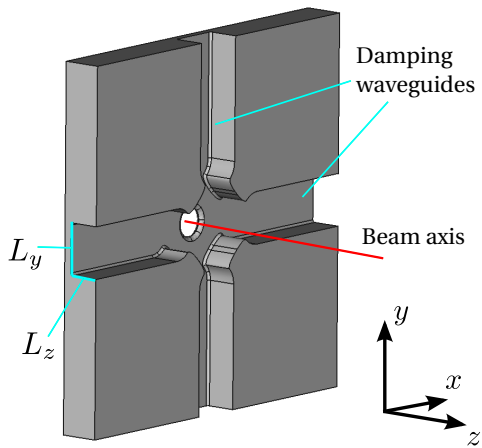}
\caption{In the above CLIC accelerating cell,
\cite{grudiev2009possible} the four radial rectangular waveguides
(terminated by electromagnetic absorbers) strongly damp
HOMs; the cutoff frequency of each waveguide is slightly above the
accelerating mode frequency and well below the lowest dipole
frequency.}
\label{fig:clicCell}
\end{figure}

In the waveguide damping technique, wakefields are reduced by coupling
HOMs out through waveguides terminated by electromagnetic absorbers. The CLIC
design includes four radially-directed waveguides
in every cell, optimized to damp the TM$_{110}$ cavity mode (since the
TM$_{110}$ mode, or lowest dipole mode, is the largest contributor to
transverse wakefields). Figure \ref{fig:clicCell} shows the second cell
of the 26-cell constant-gradient CLIC cavity, TD26\_vg1.8\_R05\_CC
\cite{grudiev2012clic} (the second cell is the first ``regular'' cell,
i.e.~the first cell
\emph{without} power input couplers).
The multicell cavity
is formed by stacking pieces similar to that shown in
Fig.~\ref{fig:clicCell}.

The waveguides of the CLIC cavity effectively damp the
lowest dipole mode (and other HOMs) {\it without} damping
the accelerating mode.
This desired behavior was obtained by carefully selecting the
waveguide cross-section such that the lowest waveguide cutoff
frequency is between the accelerating mode and (undamped) dipole mode
frequencies. For the CLIC cell in Fig.~\ref{fig:clicCell}, the lowest
waveguide cutoff frequency is 
$c / 2 L_y = 13.6$ GHz ($L_y = 11$mm),
the accelerating mode
frequency is 12 GHz, and the frequency of an undamped dipole mode
(i.e.~dipole mode in a CLIC cell {\it without} damping waveguides)
is approximately 21 GHz.
Since the frequency of the accelerating mode
is below waveguide cutoff, the mode remains confined to the cavity;
the dipole mode, however, is free to propagate down the
waveguides.
The cutoff frequency is chosen far below the undamped
dipole frequency because damping is less effective near cutoff
where the group velocity (and thus speed of energy transport)
of the waveguide mode vanishes \cite{kroll1993persistent,lin1995minimum}.

\section{Intrinsic damping in PhCs}


Photonic crystals offer an alternative approach to wakefield damping:
confine only
the accelerating mode. This approach is made possible by the bandgap
property of some PhCs, and the resulting defect-mode phenomenon.
A brief overview follows.



\begin{figure}[htbp]
\begin{center}
\includegraphics[width=0.48\textwidth]{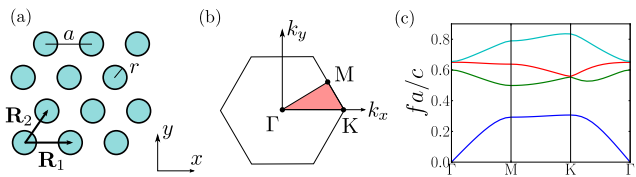}
\end{center}
\caption{Propagation in the 2D triangular lattice of sapphire discs
(a) is forbidden for TM electromagnetic waves with frequencies near
$0.4c/a$ because of the bandgap (c).  Lattice vectors ${\bf R}_1$ and
${\bf R}_2$, inter-disc spacing $a$, and disc radius $r$ are defined
in (a). The first Brillouin zone of the reciprocal lattice is
identified in (b) by the hexagon; the {\it irreducible} Brillouin zone
is shaded and represents the entire Brillouin zone by symmetry. The
dispersion curves for the first 4 bands along the path outlining the
shaded region in (b) are shown in (c) for $r = 0.17a$.  Calculations
were performed using the MIT Photonic Bands simulation code
\cite{johnson2001mpb}.}
\label{fig:trilatt}
\end{figure}

A {\it band diagram} succinctly summarizes the electromagnetic
properties of PhCs. It is well-known from Bloch theory that
electromagnetic eigenmodes in periodic structures take the form
\begin{equation}
\E(\bx, t) = \tilde{\E}_n(\bx) e^{i {\bf k} \cdot \bx - i \omega_n({\bf
k}) t}
\label{eq:blochMode}
\end{equation}
where the reciprocal lattice vector $\bf k$ runs over a reduced
region of $k$-space called the first Brillouin zone, $n$ is an
integer indexing ``bands'' of solutions, and $\tilde{\E}_n$ has the
periodicity of the lattice. A band diagram traditionally
plots the resonant
frequencies $\omega_n$ in each band along a
representative $k$-space path within the first Brillouin zone.
Figure \ref{fig:trilatt}c shows the TM band diagram for the PhC of
interest to this paper (2D triangular lattice of sapphire
discs---shown in Fig.~\ref{fig:trilatt}a)
along the $k$-space path shown in Fig.~\ref{fig:trilatt}b.

\begin{figure}[htbp]
\begin{center}
\includegraphics[width=0.5\textwidth]{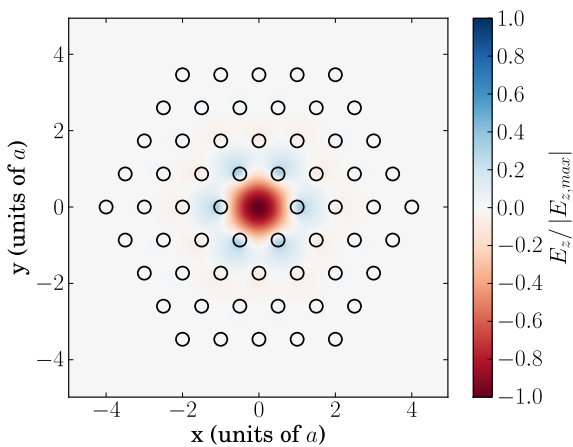}
\end{center}
\caption{Resonant TM defect cavity mode in a triangular lattice of
lossless sapphire discs ($r = 0.17a$). The lattice is truncated at 4
layers (60 discs), giving a radiative $Q$-factor of $Q_{\rm rad} =
24000$. $Q_{\rm rad}$ increases exponentially with the number of
layers.}
\label{fig:trilattcav60}
\end{figure}

Of central importance in Fig.~\ref{fig:trilatt}c
is the wide range of frequencies
between the first and second band in which there are no solutions.
This feature is called a bandgap, and is the basis for the formation
of PhC resonant cavities that are of interest to the accelerator
community. For a PhC with a bandgap, a resonant cavity can often be
formed by removing a single element from the lattice (creating a
defect). This introduces a mode that oscillates at a frequency
within the bandgap and is necessarily localized to the defect. The
defect TM mode for the sapphire triangular lattice is shown in Fig 
\ref{fig:trilattcav60} and oscillates at a frequency of
$f = 0.41 c / a$---the center of the bandgap; this
mode can accelerate particles in/out of the page.

\begin{figure}[htbp]
\begin{center}
\includegraphics[width=0.4\textwidth]{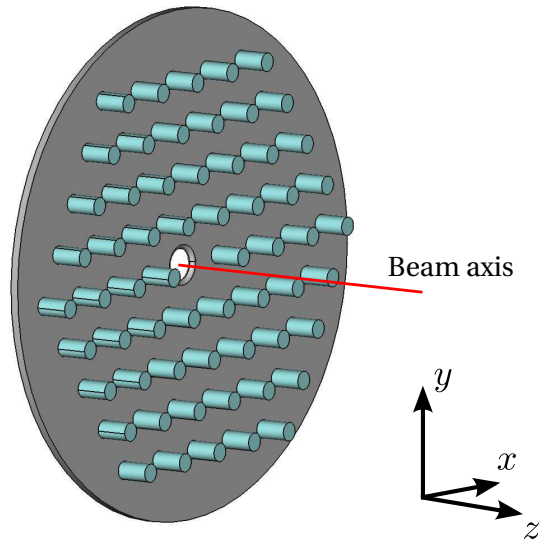}
\end{center}
\caption{The triangular lattice PhC of sapphire rods confines the
accelerating mode to the beam axis while damping HOMs.}
\label{fig:tri4Cell}
\end{figure}

A single cell of a hybrid PhC cavity (Fig.~\ref{fig:tri4Cell})
is formed from a 2D PhC of sapphire rods
(in this case a triangular lattice) sandwiched between two
copper iris plates. As with the CLIC cavity, a
multicell version would be assembled by stacking the element shown in
Fig.~\ref{fig:tri4Cell}.

The dielectric constant of sapphire is
anisotropic; in the case of Fig.~\ref{fig:tri4Cell}, the c-axis
of the sapphire is oriented along $z$. This means the dielectric
tensor used in our calculations took the form: $\varepsilon = 9.4
\varepsilon_0 (\hat{\bf x} \hat{\bf x} + \hat{\bf y}\hat{\bf y}) +
11.6 \varepsilon_0 \hat{\bf z} \hat{\bf z}$ where $\varepsilon_0$ is
the vacuum permittivity. 

This structure intrinsically damps HOMs because
frequencies above the bandgap can propagate through the crystal
(assuming no higher-frequency bandgaps).
However, as with waveguide damping, one must be wary of flat
portions of the dispersion curves, where group velocity (and
thus energy transport) vanishes \cite{li1997wake}.
In the following
results and analysis, we find that for the 2D triangular lattice PhC,
dipole resonances in the defect are damped poorly by the surrounding
PhC. The cause is the coincidence of the dipole resonant frequency
with the upper edge of the bandgap, where the second band flattens.


\section{Comparison of wakefield in CLIC and the truncated PhC cavity}
\label{sec:results}

We performed time-domain wakefield simulations for 8-cell versions of
the CLIC (Fig.~\ref{fig:clicCell}) and Tri-4-Sapphire
(Fig.~\ref{fig:tri4Cell}) cavities \cite{WilsonWakeFieldGuide}.
The 8 cells were identical
in each cavity
type (approximating the behavior of an infinitely periodic 
single cell). The iris geometry of the Tri-4-Sapphire cavity
was matched with the second cell of
the TD26\_vg1.8\_R05\_CC cavity, defined by the iris radius
$a = 3.15$mm and iris thickness $d = 1.67$mm; as a result, short-range
wakefields (those wakefields affecting the drive bunch)
are identical in both cavities.


Each cavity was driven by a highly relativistic electron bunch with
a Gaussian profile in $z$, and a delta-function profile in $x$ and
$y$. The Gaussian half-width of the bunch
was $\sigma_z = 1$mm; therefore, HOMs up to frequencies of $\approx
90$ GHz were excited significantly.
To excite transverse wakefields, the bunch was offset (from the
beam axis) by 1mm in the $x$-direction.
Wakefields were absorbed by normally conducting layers at the
simulation domain boundaries (for further discussion on the
performance of these absorbers, see Appendix \ref{app:absorb}).

All simulations were implemented in the \textsc{vorpal} framework which
uses the finite-difference time-domain method for electromagnetics and
the particle-in-cell technique for simulating the electron bunch and
test particles \cite{JComp.196.448}.
In all simulations, grid cells were cubic with
$\Delta z = 0.3$mm so that the excitation bunch cutoff frequency was
simulated with $\approx 10\%$ accuracy.

\subsection{Wake potential}

The {\it wake potential} is the net momentum change (normalized by
charge) of a point charge trailing the wakefield excitation bunch.
The {\it longitudinal} wake potential is defined as
\begin{equation}
W_z(s, \br, \brp) = - \frac{1}{q_e} \int_0^L E_z(z, \br, \brp, t =
(s + z) / v) \, dz
\end{equation}
where $s$ is the point charge's distance (along $z$)
behind the density peak of the Gaussian excitation bunch, $\br$ is the
transverse position of the point charge (the beam axis is the
transverse origin), $\brp$ is the transverse position of the
excitation bunch, $v$ is the bunch/point charge velocity (basically
$c$), $L$ is the total length of the multicell cavity,
and $q_e$ is the excitation bunch charge.
Similarly, the transverse wake potential is defined as
\begin{equation}
{\bf W}_{\perp}(s, \br, \brp) = \frac{1}{q_e} \int (
\E_{\perp} + v \hat{z} \times {\bf B}_{\perp}) (z,
\br, \brp, t = (s + z) / v) \, dz.
\end{equation}

In the theoretical case of
a cylindrically symmetric cavity with infinitely long beam tubes, the
form of the $m$th-order azimuthal multipole of the
wake potential (for $|\br| < a$) is particularly simple
\cite{napoly1993generalized}:
\begin{align}
W_{z, m}(r, r^{\prime}, \phi, s) &= \partial_s X_m(s) r^m {r^{\prime}}^m
\cos m \phi \label{eq:mpoleZWake} \\
{\bf W}_{\perp, m}(r, r^{\prime}, \phi, s) &= m X_m(s) r^{m-1} {r^{\prime}}^m
\left(\cos m \phi \, \hat{r} - \sin m \phi \, \hat{\phi} \right)
\label{eq:mpoleTWake}
\end{align}
where $\phi^{\prime}$ has been set to zero in the above (since only the
difference in azimuth matters for cylindrical symmetry) and $m > 0$
for Eq.~\ref{eq:mpoleTWake}. The transverse wake potential is zero
when $m=0$; only modes with $m > 0$ mediate transverse kicks
in this case \cite{napoly1993generalized}.

The longitudinal and transverse wake potentials are dominated by
monopole and dipole wakefields, respectively. The expressions for
these contributions are simpler still:
\begin{align}
W_{z,0} &= \partial_s X_0(s) \\
{\bf W}_{\perp,1} &= \br^{\prime} X_1(s).
\end{align}
Both expressions are uniform throughout the beam tube
region and the dipole transverse wake potential is proportional
to and in the
direction of the drive beam's transverse offset.

The structures under consideration are not cylindrically symmetric;
nevertheless, the above forms are assumed in our calculations, since
the forms hold well for wake potentials near the beam axis
\cite{napoly1993generalized,werner2009wakefields}.
We are mainly interested in the monopole and dipole
contributions (i.e.~the largest longitudinal and transverse kicks),
and therefore plot $\partial_s X_0(s) / L$ and
$X_1(s) / L$, in units of V/pC/m and V/pC/m/mm, respectively.

The wake potential was calculated by chasing the drive electron bunch with
test particles of very low charge (so that wakefields were not induced
by the test particles), but equally diminished mass to retain the electron
charge-mass ratio. The net momentum change of the test particles
indirectly gives the wake potential. The test particles were organized
in rings so that the multipole contributions could be extracted. Six
test particles were placed 1 mm from the beam axis.

\subsection{Results}

\begin{figure}[htbp]
  \centering
  \includegraphics[width=0.45\textwidth]{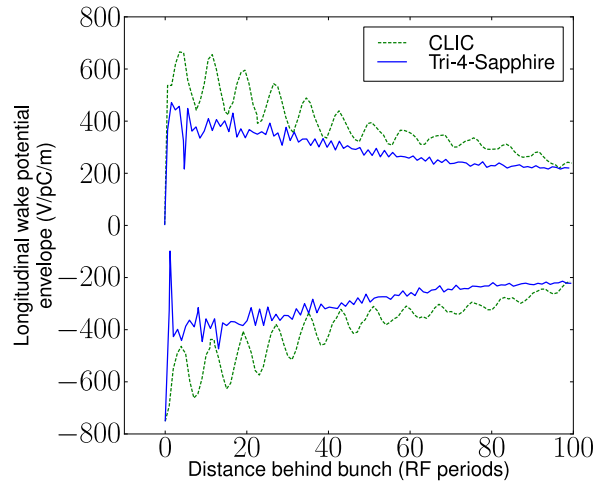}
  \caption{Oscillations in the envelope of the monopole wake potential
  indicate the presence of monopole HOMs.}
  \label{fig:z0WakeTri4}
\end{figure}

\begin{figure}[htbp]
  \centering
  \includegraphics[width=0.45\textwidth]{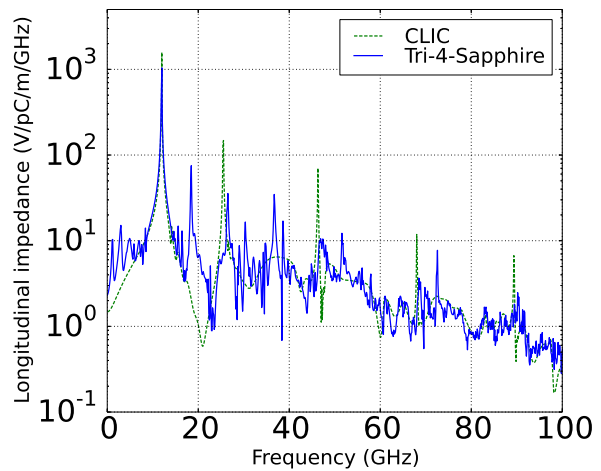}
  \caption{Monopole HOMs are relatively numerous in the Tri-4-Sapphire
  cavity, but occur at lower amplitudes.}
  \label{fig:z0ImpTri4}
\end{figure}

The envelopes (lines connecting local maxima/minima) 
of the longitudinal monopole wake potentials are shown in
Fig.~\ref{fig:z0WakeTri4} with the corresponding wake impedances shown
in Fig.~\ref{fig:z0ImpTri4}. Fluctuations of the envelope give a 
qualitative picture of the presence of HOMs (showing the full wake
potential only serves to confuse the plot).
The impedances in Fig.~\ref{fig:z0ImpTri4} show lower peaks
for the Tri-4-Sapphire cavity, indicating stronger monopole HOM
damping and/or weaker monopole HOM excitation. The mode
density is significantly higher in Tri-4-Sapphire.

%

\begin{figure}[htbp]
  \centering
  \includegraphics[width=0.45\textwidth]{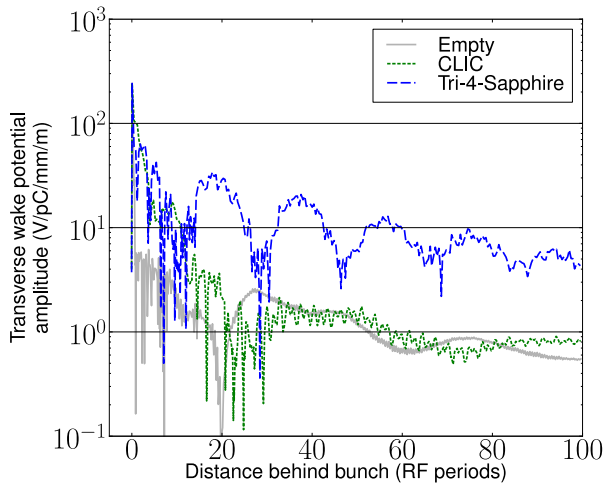}
  \caption{Transverse dipole wake potentials in the Tri-4-Sapphire
  cavity are (on average) an order
  of magnitude higher than those in CLIC.}
  \label{fig:t1WakeHex60}
\end{figure}

\begin{figure}[htbp]
  \centering
  \includegraphics[width=0.45\textwidth]{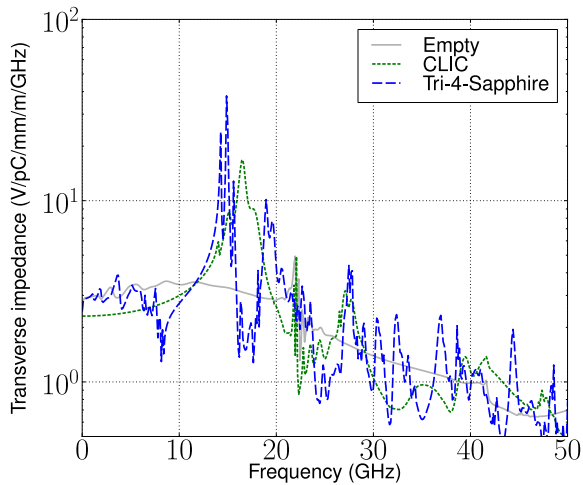}
  \caption{The higher transverse wake in the Tri-4-Sapphire cavity is
  due to the cluster of ``high''-$Q$ modes around 15 GHz.}
  \label{fig:t1ImpHex60}
\end{figure}

Amplitudes of the transverse dipole wake potentials are shown in
Fig.~\ref{fig:t1WakeHex60} with the corresponding impedances shown
in Fig.~\ref{fig:t1ImpHex60}. As a reference, we have included
in each plot
the transverse dipole wake potentials in the {\it empty}
cavity---defined most easily as an 8-cell hybrid PhC cavity without
{\it any} rods. The empty cavity represents ideal wakefield damping,
but does not confine an accelerating mode.
Dipole wakes are relatively high in the Tri-4-Sapphire cavity;
the frequencies of the troublesome HOMs are indicated by the sharp
peaks around 15 GHz in Fig.~\ref{fig:t1ImpHex60}. This result
motivates a more careful analysis of the intrinsic PhC damping
mechanism and prompts the following discussion.

\section{Origin of wakefields in the truncated PhC cavity}
\label{sec:discuss}

The transverse dipole impedance in the Tri-4-Sapphire cavity
shows that the dipole wake potential mostly comprises 
a cluster of
modes near 15 GHz. In this section, we explain the presence
of these modes using the properties of
the triangular-lattice band structure:
flat regions of the dispersion diagram imply low-group-velocity
PhC modes which sluggishly transport energy and thus cannot
effectively damp cavity modes to which they are strongly coupled.
This issue was addressed briefly in
Ref.~\cite{li1997wake} within a study of a square-lattice-based
metal-rod accelerating structure.

We have discussed
how waveguide damping is effective for coupling out frequencies only
above cutoff (the further above cutoff the better).  This is because
waveguide modes with frequencies near cutoff have vanishing group
velocities (because of the flattened dispersion at cutoff) and thus
propagate slowly down the waveguide---effectively, they are trapped.
In PhCs, the dispersion tends to flatten where the spatial variation
of the modes matches the spatial variation of the dielectric,
introducing PhC modes with vanishing group velocities. 

\begin{figure}[htbp]
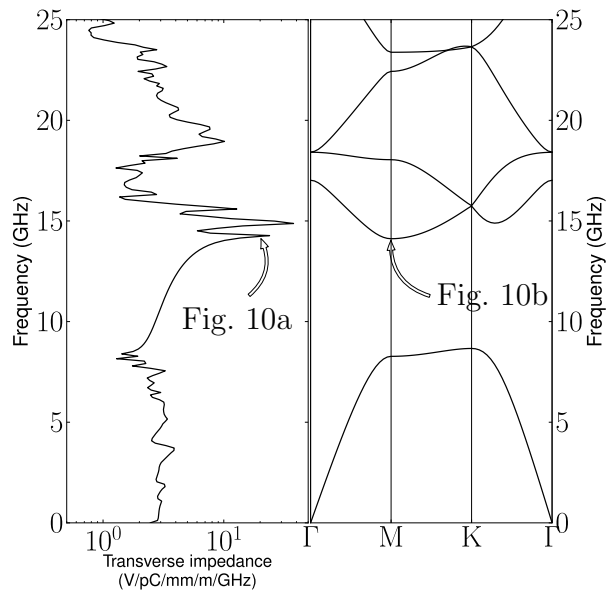

  \centering
\begin{overpic}[width=0.45\textwidth]{\figpath{tri4T1WakeVsBands}}
\put(29,45){\large Fig.~\ref{fig:tri4EzDipole}}
\put(71,49){\large Fig.~\ref{fig:ezMpointBand2}}
\end{overpic}
  \caption{The left plot shows a zoomed-in view of the troublesome
  part of the transverse dipole impedance (from
  Fig.~\ref{fig:t1ImpHex60}).
  The
  right plot is the 2D TM band diagram for the triangular lattice of
  sapphire discs. The annotated peak in the impedance is at the same
  frequency as the M-point of the second band (also see the mode
  patterns in Fig.~\ref{fig:ezDipole}). This correlation
  supports the idea that low-$v_g$ modes pose a problem for
  wakefields.}
  \label{fig:tri4T1WakeVsBands}
\end{figure}

\begin{figure}[htbp]
  \centering
  \subfloat[Tri-4-Sapphire mode]{
    \label{fig:tri4EzDipole}
    \includegraphics[width=0.45\textwidth]{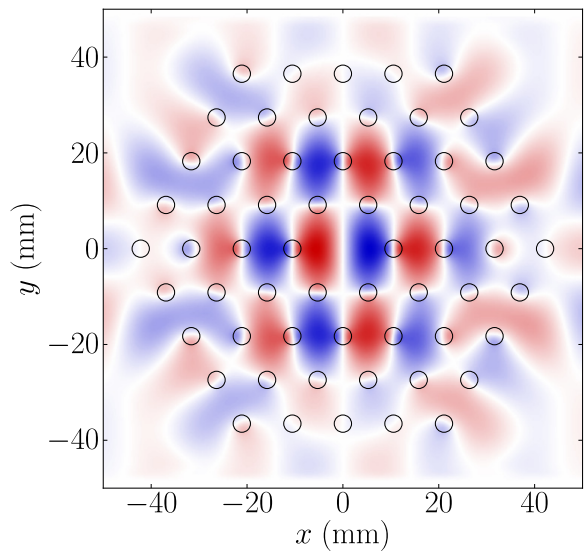}
  } \\
  \subfloat[Infinite PhC mode]{
    \label{fig:ezMpointBand2}
    \includegraphics[width=0.45\textwidth]{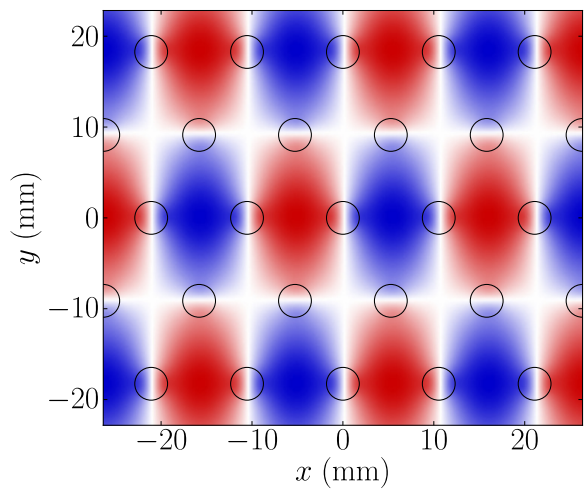}
  }
  \caption{Analagous modes in (a) the Tri-4-Sapphire periodic single-cell
  cavity and (b) an infinite 2D PhC
  (the mode pattern in (b)
  is uniform in $z$). The propagating PhC mode in (b) has a vanishing group
  velocity; thus, its counterpart in (a) is ``trapped'' and
  contributes significantly to the dipole wakefield. Computations used (a)
\textsc{vorpal} \cite{JComp.196.448} and (b) MIT Photonic Bands
\cite{johnson2001mpb}.}
  \label{fig:ezDipole}
\end{figure}

Figure \ref{fig:tri4T1WakeVsBands} shows the transverse dipole
impedance of the Tri-4-Sapphire cavity in the frequency range 0-25 GHz
(a zoom view of Fig.~\ref{fig:t1ImpHex60})
and matches it with the 2D TM band diagram for the triangular lattice
of sapphire discs. The two most prominent peaks in the impedance
clearly line up with the flat portions of the second band (the third
peak can be matched with another flat band in the fully 3D band
diagram). The annotated impedance peak (second largest) was
investigated further using the time-domain mode extraction technique.
Figure \ref{fig:ezDipole} shows the field pattern of the extracted
mode (from a periodic single-cell simulation at phase advance $\phi
\approx 3 \pi / 4$) and compares it with the 2D PhC lattice mode at
the M-point of the second band. The field patterns clearly match.


Both the damping waveguides and the triangular PhC suffer from
low-group-velocity modes. Why then is the damping worse in the
Tri-4-Sapphire cavity? The answer lies in the combined characteristics
of the damping structure and the central cavity region.  Consider the
lowest dipole mode in the CLIC cavity. Section \ref{sec:clicWg}
showed that waveguide damping is more effective when the undamped
resonant frequency is further above the cutoff frequency. The central
cavity dimensions in the CLIC cavity determine a dipole resonant
frequency; thus, the waveguide dimensions are selected such that
cutoff is as far below that frequency as possible (without affecting
the accelerating mode too much). The gap between the accelerating
frequency and the dipole frequency is large enough such that this
scheme results in effective dipole damping.

In contrast, the defect region of the Tri-4-Sapphire cavity is (by
definition) highly commensurate with the geometry of the surrounding
lattice.  Furthermore, the mode patterns in Fig.~\ref{fig:ezDipole}
show little electric field energy inside the dielectric, indicating that the
removal of the central rod has a small, perturbative effect (compared
to modes at the top of the first band, which have most of their field
energy concentrated in dielectric). (For an introduction to
electromagnetic perturbation theory, see
Ref.~\cite{joannopoulos2011photonic}.)
As a result, the dipolar defect resonance is at the same frequency as
the low-group-velocity lattice mode (or the ``cutoff'' mode), and is thus
poorly damped.  Put another way: the creation of the defect
weakly perturbs the band-edge lattice mode; therefore, the mode retains a
strong presence in the defect. This argument suggests pushing the
inner layer of rods closer to the beam axis, which is likely to
increase the dipole resonance frequency in the defect above the flat
portion of the lattice band and thus increase coupling to
higher-group-velocity lattice modes.

The situation may be exacerbated by the impedance mismatch at the
outer layer of rods. Because of its low-group-velocity, the lattice
mode shown in Fig.~\ref{fig:ezDipole} may be highly susceptible to
reflections off of the transition between lattice and vacuum,
effectively increasing its confinement. This transition could be
made smoother by slowly decreasing the radii of the rods in outer
layers, but would add significantly to the transverse size of the
structure.

\section{Wakefields of optimized hybrid PhC cavities}
\label{sec:opt}

\begin{figure}[htbp]
\begin{center}
\includegraphics[width=0.4\textwidth]{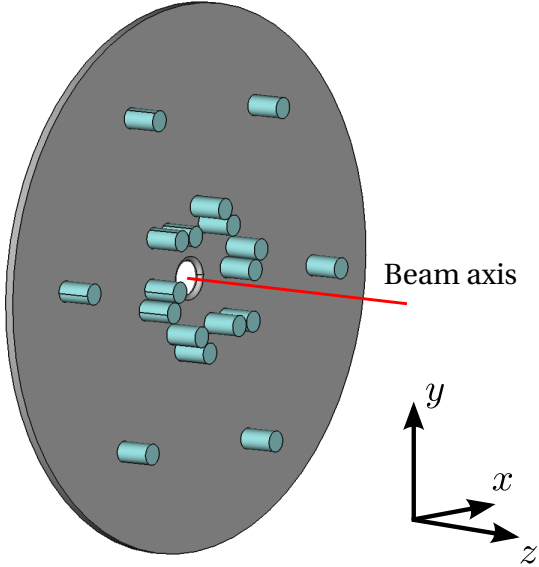}
\end{center}
\caption{An optimized hybrid PhC cavity (Opt-18-Sapphire)
reduces the number of rods required to confine the accelerating mode.}
\label{fig:opt18}
\end{figure}

In a previous work, 2D simulations showed that some rod arrangements lacking
any lattice symmetry (but retaining some rotational symmetry)
dramatically reduce the radiative losses of the accelerating mode as
compared to lattice arrangements of equal rod count
\cite{bauer2008truncated}. In this section, we construct and simulate
a 3D multicell cavity from a 2D optimized structure by adding
conducting iris plates as done with the triangular lattice PhC. A
single cell is pictured in Fig.~\ref{fig:opt18}
and will be referred to as Opt-18-Sapphire (rods are sapphire).

The 2D optimizations in Ref.~\cite{bauer2008truncated} ignored the
effect of irises; ideally, the 3D cavity would be re-optimized in
their presence. The Opt-18-Sapphire cavity in this work has \emph{not}
been re-optimized in 3D; such a pursuit is left to future work. Based
on the sensitivity study in Ref.~\cite{bauer2008truncated}, we
expect only minor rod displacements (from
the 2D optimized arrangement) in 3D optimizations given a small
perturbative effect due to the irises.
See Appendix \ref{app:accMode} for further discussion of the
accelerating mode in Opt-18-Sapphire.

%

\begin{figure}[htbp]
  \centering
  \includegraphics[width=0.45\textwidth]{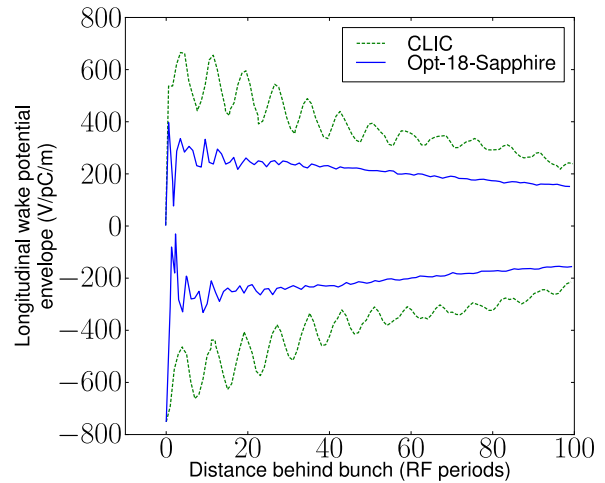}
  \caption{The monopole wake potential envelope is smoother for the
  Opt-18-Sapphire cavity. However, the amplitude of the accelerating
  mode is lower, indicating weaker coupling to the beam
  (see Appendix \ref{app:accMode}).}
  \label{fig:z0WakeOpt18}
\end{figure}

\begin{figure}[htbp]
  \centering
  \includegraphics[width=0.45\textwidth]{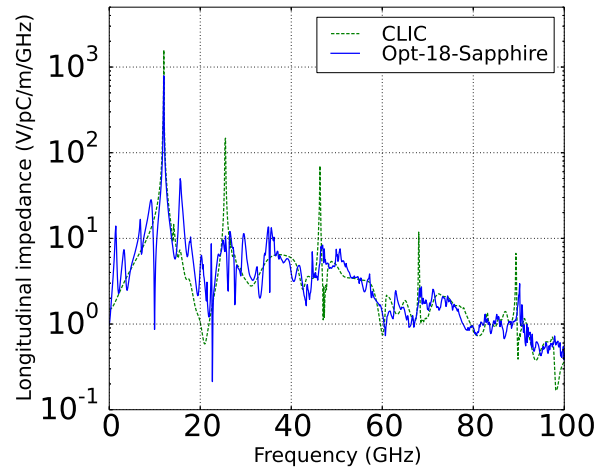}
  \caption{Monopole HOMs are numerous but suppressed in the
  Opt-18-Sapphire cavity.}
  \label{fig:z0ImpOpt18}
\end{figure}

\begin{figure}[htbp]
  \centering
  \includegraphics[width=0.45\textwidth]{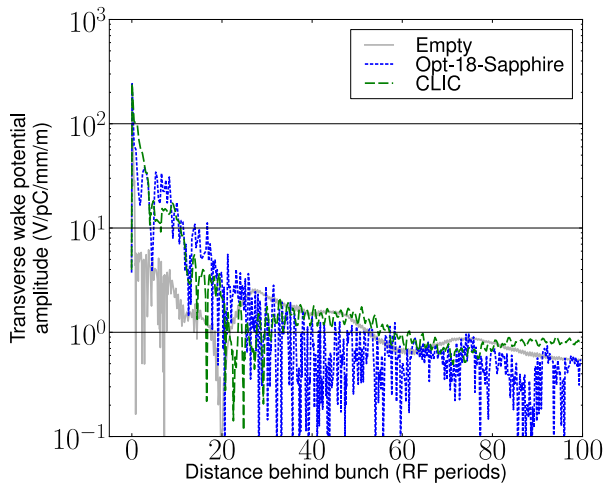}
  \caption{Optimized 18-rod cavity reduces transverse wakes
  Transverse dipole wake potentials in 8-cell cavities using
  conducting absorbers.}
  \label{fig:t1WakeOpt18}
\end{figure}

\begin{figure}[htbp]
  \centering
  \includegraphics[width=0.45\textwidth]{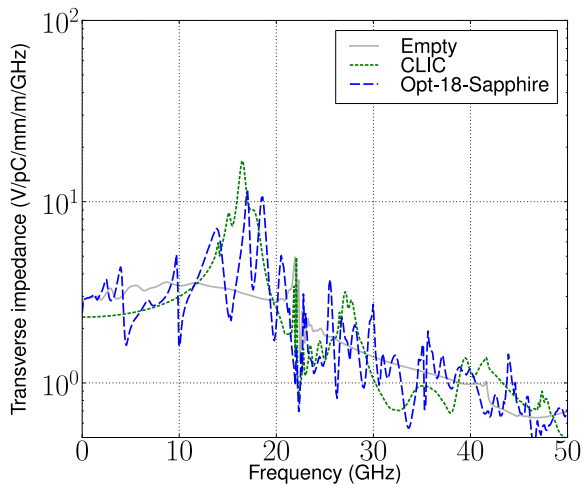}
  \caption{Transverse dipole impedance in 8-cell cavities using
  conducting absorbers.}
  \label{fig:t1ImpOpt18}
\end{figure}

Simulations in Ref.~\cite{werner2009wakefields} showed 
that optimized structures can reduce wakefields; however,
the simulations in that study were of single-cell closed cavities (no
irises or beam tubes) and made comparisons with only the pillbox.
Figures \ref{fig:z0WakeOpt18}--\ref{fig:t1ImpOpt18} detail the wake
potentials of an 8-cell Opt-18-Sapphire cavity and compare with CLIC.

Monopole HOMs are more numerous but less prominent in the
Opt-18-Sapphire cavity. Transverse wakes in Opt-18-Sapphire
are similar to CLIC on average, but \emph{greatly} reduced
in comparison with Tri-4-Sapphire (cf.~Fig.~\ref{fig:t1WakeHex60}).
Unfortunately, the analysis of Section \ref{sec:discuss} cannot be
applied to Opt-18-Sapphire because of the lack of any band structure;
however, these results suggest that non-lattice structures could
avoid the inherently poor damping characteristics found in PhC defect
cavities.

\section{Concluding remarks}

Low wakefields require that the energy in HOMs be dissipated as
quickly as possible. Thus, a damping mechanism (or structure) should
have a mode spectrum without any low group velocities.  In principle,
the CLIC cavity suffers from low-$v_g$ HOMs near the cutoff frequency
of its damping waveguides; however, the sparse mode density of the
conducting cavity allowed the placement of the cutoff frequency within
an empty region of the spectrum, producing effective damping. PhC cavities
suffer from low-$v_g$-confined HOMs due
to the flattening of bands near $k$-points of strong lattice symmetry.
Specifically, in the Tri-4-Sapphire design, the defect-based cavity
supports dipolar resonances that strongly couple to band-edge PhC
modes with low-$v_g$.

Future work should focus on reducing wakefields in lattice-based PhC
cavities because the lattice structure has the potential to reduce
surface magnetic fields and increase accelerating efficiencies (see
Appendix \ref{app:accMode}) and PhC band theory provides an
explanation for the existence of troublesome HOMs.
Possible routes to reducing
wakes in Tri-4-Sapphire cavities include perturbing the central defect
region to eliminate low-$v_g$ resonances and/or reducing impedance
mismatch at the truncation of the lattice. Given the low wakefields
in the Opt-18-Sapphire cavity (and its high $Q_{\rm rad}$ to rod-count
ratio), brute-force optimization of rod positions to lower wakefields 
is also recommended.

\section{Acknowledgments}

This work was supported by the U.S. Department of Energy grant
DE-FG02-04ER41317.

Most of the simulations described in this work were performed with the
\textsc{Vorpal} framework; we
would like to acknowledge the efforts of the \textsc{Vorpal} team:
D.~Alexander,
K.~Amyx,
E.~Angle,
T.~Austin,
G.~I.~Bell,
D.~L.~Bruhwiler,
E.~Cormier-Michel,
Y.~Choi,
B.~M.~Cowan,
R.~K.~Crockett
D.~A.~Dimitrov,
M.~Durant,
B.~Jamroz,
M.~Koch,
S.~E.~Kruger,
A.~Likhanskii,
M.~C.~Lin,
M.~Loh,
J.~Loverich,
S.~Mahalingam,
P.~J.~Mullowney,
C.~Nieter,
K.~Paul,
I.~Pogorelov,
C.~Roark,
B.~T.~Schwartz,
S.~W.~Sides,
D.~N.~Smithe,
P.~H.~Stoltz,
S.~A.~Veitzer,
D.~J.~Wade-Stein,
N.~Xiang,
C.~D.~Zhou.

We would also like to thank Alexej Grudiev at CERN for providing the
CLIC cavity geometry.

\appendix

\section{Absorbers}
\label{app:absorb}


To absorb wakefields, the ends of each damping waveguide and
the transverse edges
of the PhC simulations were filled with blocks of conducting
material with a special conductivity profile aimed at minimizing
reflections off of the cavity-facing surface. The profile is given by
\begin{equation}
\sigma(w) = \sigma_{\rm max} \left(\frac{w}{W}\right)^2
\end{equation}
where $w$ is depth in the conductor
and $W$ is the entire depth of the conducting block. The quadratic form
resembles damping techniques used in practice \cite{luong1999rf},
where a cone of
absorbing material is placed at the end of a damping waveguide. Our
calculations used $W = 25$mm and $\sigma_{\rm max}/\eps_0 = 4.7
\omega_{\rm acc}$ where $\omega_{\rm acc} = 2 \pi 12$ GHz.

\begin{figure}[h]
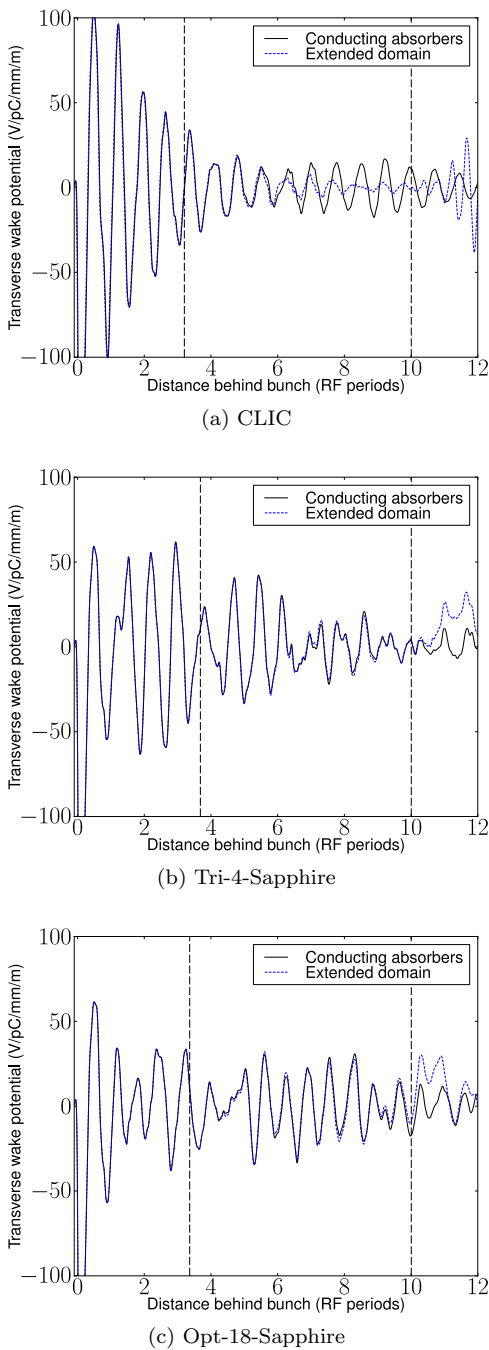

  \centering
  \subfloat[CLIC]{
    \label{fig:clicAbsCompare}
    \includegraphics[width=\abspicwidth]{\figpath{clicNoAbsVsCondAbs}}
  } \\
  \subfloat[Tri-4-Sapphire]{
    \label{fig:tri4SapphAbsCompare}
    \includegraphics[width=\abspicwidth]{\figpath{tri4NoAbsVsCondAbs}}
  } \\
  \subfloat[Opt-18-Sapphire]{
    \label{fig:opt18SapphAbsCompare}
    \includegraphics[width=\abspicwidth]{\figpath{opt18NoAbsVsCondAbs}}
  }
  \caption{Reflections off of the conducting absorbers
cause differences in the wake potentials between the two vertical
lines (more so in the CLIC cavity).
Beyond the second line, reflections from the extended domain boundaries
reach the beam axis.}
  \label{fig:absCompare}
\end{figure}

We have found that the additional contribution to the wake potentials
due to
reflections off of the conducting absorbers are higher in the CLIC
cavity than in the hybrid PhC cavities. These findings were obtained from
simulations with large transverse extents---125mm from beam axis to
simulation edge in both $x$ and $y$ (simulations in Section
\ref{sec:results} had approximately 40mm between the beam axis and the
start of the absorbers). 
Wakefield reflections
off of the absorbers at 40mm show up in the wake potential at $s
\gtrsim 80$mm, whereas reflections from 125mm absorbers show up at $s
\gtrsim 250$mm. Transverse dipole wake potential results
were compared at distances of $80$mm $< s < 250$mm (see
Fig.~\ref{fig:absCompare}). Differences in
this
region were significant for the CLIC cavity and minor in all
hybrid PhC cavities.
This means wakefield damping can
be improved by reducing the impedance mismatch of the absorber in the
CLIC waveguides, whereas improving absorption in the PhC cavities will
have little impact on the wake potentials.

\section{Accelerating mode comparison}
\label{app:accMode}

Accelerator cavity design is a complex multidimensional optimization
process that tries (in no particular order)
(1) to maximize the accelerating electric field (or gradient) so
that high energies are reached over short distances, (2) to maximize the
transfer of electromagnetic power to the beam, and (3) to minimize
wakefields to avoid beam instabilities and breakup.
The main body of this paper focused on wakefields in hybrid PhC
cavities (i.e.~goal (3)). This Appendix addresses some aspects of
goals (1) and (2) so that a more complete comparison between the CLIC
and hybrid PhC cavities may be drawn.


Using \textsc{vorpal}, figures of merit for the accelerating mode
in each cavity type were calculated and compared for a
periodic single cell with iris radius $a = 3.15$mm, iris thickness
$d = 1.67$mm, and phase advance $\phi = 2 \pi / 3$
and are summarized in Table \ref{tab:phcFom3d}. The quality
factors quantify power losses due to various physical mechanisms;
\begin{itemize}
\item $Q_{\rm rad}$: Losses due to radiation
leaking through the
truncated PhC structures (confinement of the accelerating mode is not
perfect). Applies to hybrid PhC cavities only.
\item $Q_{\rm metal}$: Losses due to RF heating of copper
surfaces.
\item $Q_{\rm diel}$: Losses due to RF heating of sapphire rods.
Applies to hybrid PhC cavities only.
We use a conservative value for the sapphire loss tangent: $10^{-4}$.
\end{itemize}

The shunt impedance per unit length is
\begin{equation}
r_{\rm shunt} = \frac{V^2}{P L}
\end{equation}
where $V$ is the accelerating voltage across a single cell,
$P$ is the total power loss per cell,
and $L$ is the length of a single cell ($L = 8.33$mm based on the
synchronicity condition). The accelerating gradient is
simply $E_{\rm acc} = V / L$. The loss factor per unit length
is $k = V^2 / 4 U L$ where
$U$ is the accelerating mode stored energy. The group velocity of the
accelerating mode is $v_g$, obtained from two simulations at different
phase advances.

\begin{table*}[hb]
\centering
\caption{Figures of merit for the accelerating mode in
periodic single cell cavities. $a = 3.15$mm, $d = 1.67$mm, $\phi =
2\pi/3$. All simulations performed at $\Delta z / d = 8$.}
\newcolumntype{R}{>{\raggedleft\arraybackslash}X}
\begin{tabularx}{\textwidth}{RRRRR} 
\hline
& Pillbox & CLIC & Tri-4-Sapphire & Opt-18-Sapphire \\ \hline \hline
$v_g / c$ (\%) & 1.83 & 1.65 & 1.16 & 0.78 \\ \hline
$Q_{\rm metal}$ & 6,700 & 5,900 & 11,400 & 11,400 \\ \hline
$Q_{\rm rad}$ & $\infty$ & $\infty$ & 26,600 & 3,800 \\ \hline
$Q_{\rm diel}$ & $\infty$ & $\infty$ & 67,000 & 39,000 \\ \hline
$Q_{\rm total}$ & 6,700 & 5,900 & 7,100 & 2,700 \\ \hline
$r_{\rm shunt}$ (M$\Omega$/m) & 106 & 82 & 70 & 18 \\ \hline
$k$ (V/pC/m) & 298 & 260 & 187 & 125 \\ \hline
$E_{\rm surf, metal, max}/E_{\rm acc}$ & 1.93 & 1.96 & 1.93 & 1.93 \\
\hline
$cB_{\rm surf, metal, max}/E_{\rm acc}$ & 1.0 & 1.54 & 1.49 & 1.73 \\
\hline
$E_{\rm surf, diel, max}/E_{\rm acc}$ & --- & --- & 0.54 & 0.64 \\
\hline
$cB_{\rm surf, diel, max}/E_{\rm acc}$ & --- & --- & 1.26 & 1.79 \\
\hline
$E_{\rm diel, max}/E_{\rm acc}$ & --- & --- & 0.60 & 0.79 \\
\hline
\end{tabularx}
\label{tab:phcFom3d}
\end{table*}



\subsection{Surface fields}

\begin{figure*}[h]
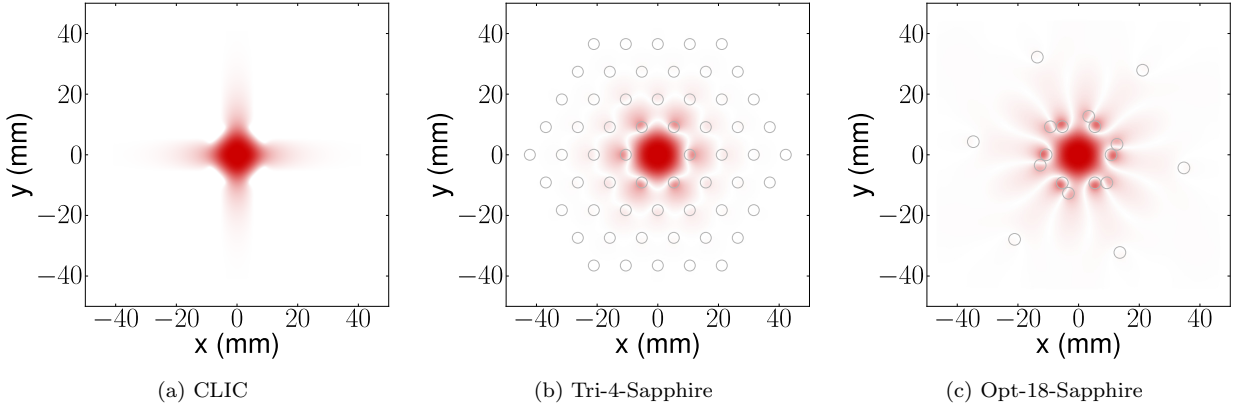

  \centering
  \subfloat[CLIC]{
    \label{fig:clicEzXY}
    \includegraphics[width=\modepicwidth]{\figpath{clicEzXY}}
  }
  \subfloat[Tri-4-Sapphire]{
    \label{fig:tri4SapphEzXY}
    \includegraphics[width=\modepicwidth]{\figpath{tri4SapphEzXY}}
  }
  \subfloat[Opt-18-Sapphire]{
    \label{fig:opt18SapphEzXY}
    \includegraphics[width=\modepicwidth]{\figpath{opt18SapphEzXY}}
  }
  \caption{Absolute value of $E_z$ on the midplane (in $z$) of
  periodic single cell cavities.}
  \label{fig:ezCompare}
\end{figure*}

\begin{figure*}[h]
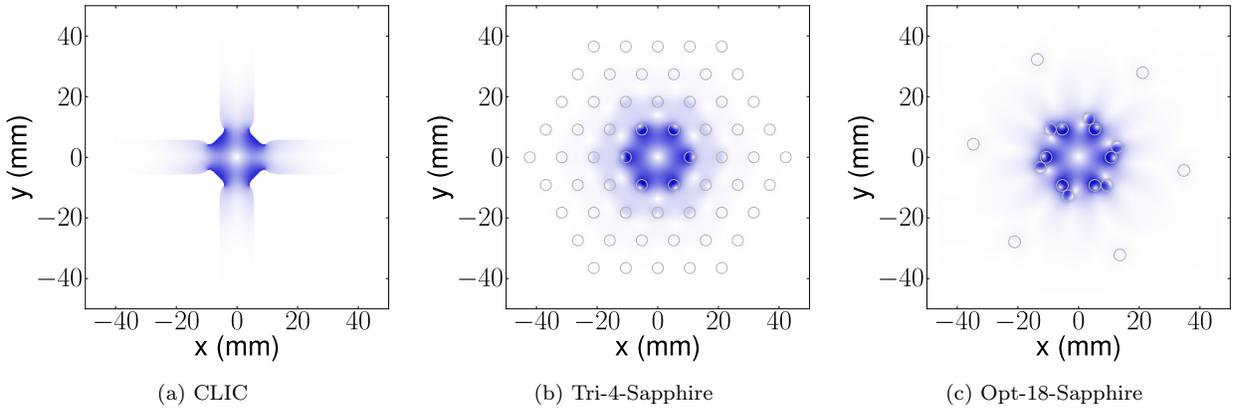

  \centering
  \subfloat[CLIC]{
    \label{fig:clicBtXY}
    \includegraphics[width=\modepicwidth]{\figpath{clicBtXY}}
  }
  \subfloat[Tri-4-Sapphire]{
    \label{fig:tri4SapphBtXY}
    \includegraphics[width=\modepicwidth]{\figpath{tri4SapphBtXY}}
  }
  \subfloat[Opt-18-Sapphire]{
    \label{fig:opt18SapphBtXY}
    \includegraphics[width=\modepicwidth]{\figpath{opt18SapphBtXY}}
  }
  \caption{Absolute value of ${\bf B}_{\perp}$ on the midplane (in $z$) of
  periodic single cell cavities.}
  \label{fig:btCompare}
\end{figure*}

Figures \ref{fig:ezCompare} and \ref{fig:btCompare} show the absolute
values of $E_z$ and ${\bf B}_{\perp}$ on the $z$-midplane of the
relevant periodic single cell cavities. The $z$-dependences of the
fields are very similar amongst the different cavity types, as
expected based on the common iris geometry in each structure.  The
maximum surface electric field occurs on the iris in all cases,
explaining the uniformity of $E_{\rm metal,surf,max} / E_{\rm acc}$
across all cavity types.  The maximum surface magnetic field occurs on
the innermost radial walls of the CLIC cavity; the elliptic curvature
of this feature was carefully chosen to minimize $c B_{\rm surf} /
E_{\rm acc}$.

In the hybrid PhC cavities, $B_{\rm metal,surf,max}$ occurs where the
innermost rods abut the conducting endplates.  Since the maximum
occurs at the interface between dielectric and conductor, the method
used to hold the rods in place will require careful consideration. For
example, brazing material could be more prone to breakdown or suffer
greater heating losses than copper.
The experiment in Ref.~\cite{masullo2006study} successfully secured the
rods simply by endplate pressure. On the other hand, covering
the region of conductor suffering the maximum magnetic field with
dielectric could suppress the breakdown mechanism.  Of course, only
experiment will tell.

\subsection{Power losses/Accelerating efficiency}

The reduced copper surface area in the hybrid PhC cavities lowers
copper heating losses compared to CLIC and the pillbox; however,
(unlike the all-copper cavities) dielectric heating and radiative
losses play a role. The shunt impedance is a strong indicator of
accelerating efficiency, and is higher in the CLIC cavity. However,
\begin{table*}[ht]
\centering
\caption{Hypothetical accelerating mode figures of merit for the
Tri-6-Sapphire cavity and the Opt-18-Sapphire cavity (given that
3D optimizations reproduce 2D $Q_{\rm rad}$ values).
Values in Table \ref{tab:phcFom3d} not
appearing here are assumed to be the same.}
\newcolumntype{R}{>{\raggedleft\arraybackslash}X}
\begin{tabularx}{\textwidth}{RRRRR} 
\hline
 & Pillbox & CLIC & Tri-6-Sapphire & Opt-18-Sapphire \\ \hline \hline
Q$_{\rm metal}$ & 6,700 & 5,900 & 11,400 & 11,400 \\ \hline
Q$_{\rm rad}$ & $\infty$ & $\infty$ & 2,400,000 & 25,000 \\ \hline
Q$_{\rm diel}$ & $\infty$ & $\infty$ & 67,000 & 39,000 \\ \hline
Q$_{\rm total}$ & 6,700 & 5,900 & 9,700 & 6,500 \\ \hline
$r_{\rm shunt}$ (M$\Omega$/m) & 106 & 82 & 97 & 43 \\ \hline
\end{tabularx}
\label{tab:phcFom3d2}
\end{table*}
shunt impedances in the hybrid PhC cavities could be increased by
reducing radiative losses (e.g.~by adding more layers of rods to the
Tri-4-Sapphire cavity, or performing optimizations in 3D on the
Opt-18-Sapphire cavity as discussed below).

In the optimized cavity, $Q_{\rm rad}$ is very sensitive to structural
perturbations (cf.~Ref.~\cite{bauer2008truncated}). We find that the
beam tubes in the 3D Opt-18-Sapphire cavity drop $Q_{\rm rad}$ by
nearly an order of magnitude relative to the 2D value
reported in Ref.~\cite{bauer2008truncated}
(from 25,000 in 2D to 3,800 in 3D with beam tubes).
It remains to be seen whether further optimization in 3D can regain
the original 2D $Q_{\rm rad}$. The $Q_{\rm rad}$ of the
Tri-4-Sapphire cavity is the same in 2D, indicating some robustness of
the confinement to structural perturbations.

Table \ref{tab:phcFom3d2} shows projected figures of merit under the
following assumptions: (1) a 3D 
optimization of the Opt-18-Sapphire cavity
(involving only minor repositioning of the rods)
restores the 2D $Q_{\rm rad}$ (25,000),
(2) two layers of rods are added
to the Tri-4-Sapphire cavity (making Tri-6-Sapphire) basically
eliminating radiative losses, and (3) the
stored energy, surface fields, and voltage gain remain the same in
each (a safe assumption if the fields are only slightly perturbed).
Of the cavities with
wakefield damping, this gives the Tri-6-Sapphire cavity the largest
shunt impedance; however,
increasing the number of lattice layers could increase wakefields.

\bibliography{phcVsClic}

%
%
%
%
%
%
%
%
%
%

\end{document}